\documentclass{sig-alternate-05-2015}

\pdfoutput=1

\usepackage{etex}
\usepackage{times}
\usepackage{booktabs} %
\usepackage[usenames, dvipsnames]{color}
\usepackage{xcolor}
\usepackage{balance}
\usepackage{url}
\usepackage{graphicx}
\usepackage{array}
\usepackage{multirow}
\usepackage{mathtools,amssymb}
\usepackage[ruled,vlined,linesnumbered]{algorithm2e}
\usepackage{verbatim}
\usepackage{comment}
\usepackage[normalem]{ulem}
\usepackage{enumitem}
\usepackage{booktabs}
\usepackage{hyperref}
\usepackage{soul}
\usepackage{multirow}
\usepackage{multicol}
\usepackage{epsfig}
\usepackage{endnotes}
\usepackage{subfig}
\usepackage{microtype}
\usepackage{enumitem}

\renewcommand{\footnotesize}{\small}

\def\Snospace~{\S{}}
\def\Fnospace~{\mbox{Fig.\hspace{0.25em}}}
\def\Tnospace~{\mbox{Tab.\hspace{0.25em}}}
\def\Enospace~{\mbox{Equation\hspace{0.25em}}}

\newcommand{\ie}{i.e.,\ }
\newcommand{\eg}{e.g.,\ }

\newcommand{\tinyskip}{\vspace{1pt}}
\newcommand{\mypar}[1]{\tinyskip\tinyskip\noindent\textbf{#1.}\xspace}

\definecolor{lightgray}{rgb}{.85,.85,.85}  %
\definecolor{orange}{RGB}{255,127,0}
\sethlcolor{lightgray} %

\newcommand{\eat}[1]{}

\def\compactify{\noitemsep \itemsep=0pt \topsep=0pt \partopsep=0pt \parsep=0pt}
\let\latexusecounter=\usecounter

\newcommand{\sysname}{Raven\xspace}
\newcommand{\raven}{\sysname}

\newcommand{\mysim}{\raise.17ex\hbox{$\scriptstyle\sim$}}

\usepackage[backend=bibtex,style=trad-abbrv,maxbibnames=1]{biblatex}
\bibliography{references}
\renewcommand*{\bibfont}{\small}
\begin{document}

\title{Extending Relational Query Processing with ML Inference}

\newcommand{\autspace}{~~~~}

\numberofauthors{13}
\author{
Konstantinos Karanasos$^1$ \autspace Matteo Interlandi$^1$ \autspace Doris Xin$^2$\thanks{The work was done while the author was at Microsoft.} \autspace Fotis Psallidas$^1$
\and
Rathijit Sen$^1$ \autspace Kwanghyun Park$^1$ \autspace Ivan Popivanov$^1$ \autspace Supun Nakandal$^3$$^*$ \autspace Subru Krishnan$^1$
\and
\autspace Markus Weimer$^1$ \autspace Yuan Yu$^1$ \autspace Raghu Ramakrishnan$^1$ \autspace Carlo Curino$^1$
\and
$^1$\textsf{Microsoft} \autspace $^2$\textsf{University of California, Berkeley} \autspace $^3$\textsf{University of California, San Diego}
}

\maketitle
\begin{abstract}	

The broadening adoption of machine learning in the enterprise is increasing the pressure for strict governance and cost-effective performance, in particular for the common and consequential steps of model storage and inference. 

The RDBMS provides a natural starting point, given its mature infrastructure for fast data access and processing, along with support for enterprise features (e.g., encryption, auditing, high-availability). To take advantage of all of the above, we need to address a key concern: {\em Can in-RDBMS scoring of ML models match (outperform?) the performance of dedicated frameworks?} 

We answer the above positively by building \emph{\raven}, a system that leverages native integration of ML runtimes (i.e., ONNX Runtime) deep within SQL Server, and  a unified intermediate representation (IR) to enable advanced cross-optimizations between ML and DB operators. In this optimization space, we discover the most exciting research opportunities that combine DB/Compiler/ML thinking. Our initial evaluation on real data demonstrates performance gains of up to $5.5\times$   from the native integration of ML in SQL Server, and up to $24\times$ from cross-optimizations---we will demonstrate \sysname live during the conference talk.

\end{abstract}

\section{Introduction}
\label{sec:intro}

Advances in machine learning (ML), first proven in high-value web applications, are fueling a trend towards digitally transforming almost every industry---in large part due to excitement around using ML to complement traditional data analysis, discover new insights, and amplify weak signals.

However, safely and effectively adopting ML in enterprise settings comes with many new challenges across model training, tracking, deployment, and inference.  We consider all those as part of our broader research agenda~\cite{flockCIDR}, but focus this paper on model inference. 

As more and more data is analyzed and monetized, concerns about securing sensitive data and risks to individual privacy have been growing considerably~\cite{gdpr}---this extends to ML models. In fact, based on interactions with enterprise customers, we expect that {\em storage and inference of ML models will be subject to the same scrutiny and performance requirements of sensitive/mission-critical operational data}.

When it comes to data, database management systems (DBMSs) have been the trusted repositories for the enterprise, as they provide fast data access and processing, as well as a mature infrastructure that delivers features such as rich connectivity, transactions, versioning, security, auditing, high-availability, and application/tool integration.
We thus propose to store and serve ML models from within the DBMS in order to extend the above described guarantees to models as well as data. However, given the current rudimentary support for ML within DBMSs, a key concern is to do so with no detriment to inference performance. 
This leads us to the key question we investigate in this paper:  {\em Can in-RDBMS scoring of ML models match (outperform?) the performance of dedicated frameworks?} 

In parallel, an interesting trend has emerged with respect to {\em inference} of ML models.
Most widely studied or promising model families can be uniformly represented~\cite{mlflowmodel}, and given a particular model, we can express how to score it on a given input using an appropriate algebra~\cite{onnx,xla}. These algebraic structures can then be executed on different environments and hardware~\cite{tensorflow,tvm,ort,pytorch}.
Among these efforts, ONNX is worth mentioning as a recent attempt for an open format to standardize ML model representation in an engine-agnostic manner, similar to the role of relational algebra in RDBMSs.
Taken together, these observations suggest that we need to consider \emph{how to incorporate ML scoring as a foundational extension of relational algebra, and an integral part of SQL query optimizers and runtimes}.

\begin{figure*}[t]
	\centering\includegraphics[width=\textwidth]{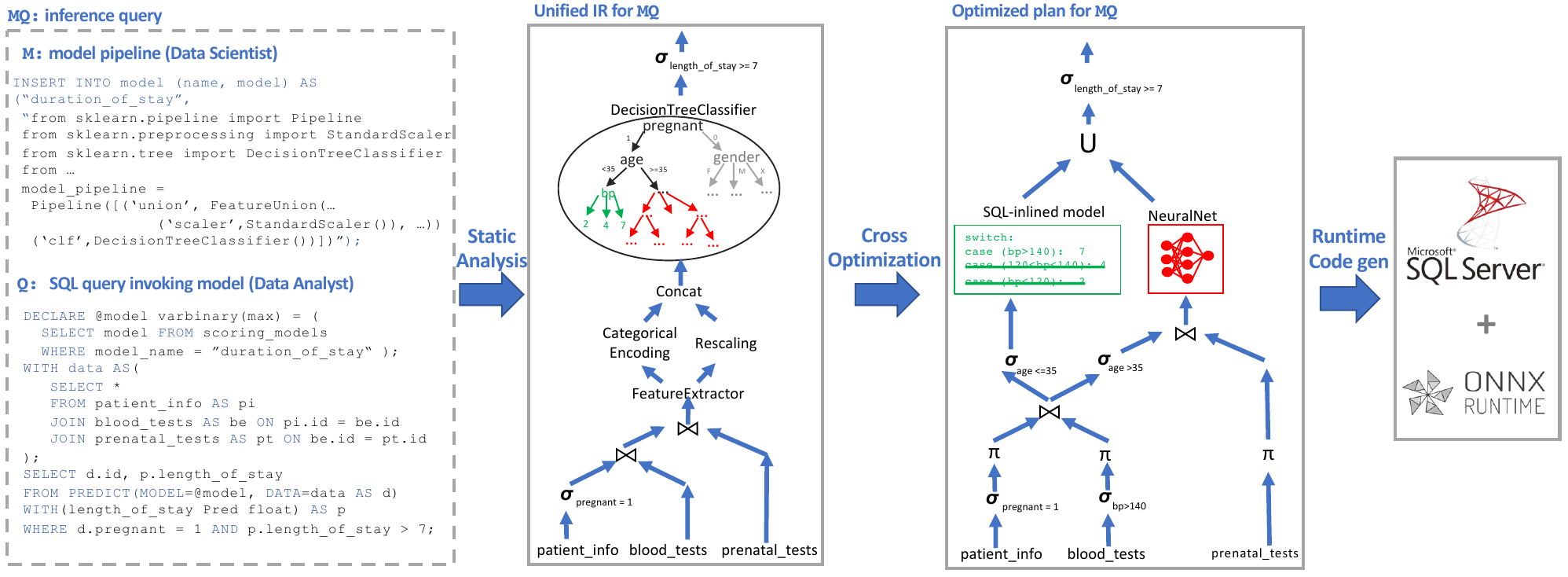}
	\vspace{-1mm}
	\caption{Running example: find pregnant patients with predicted length of stay in the hospital longer than a week.}
    \label{fig:runningexample}
    \vspace{-2mm}
\end{figure*}

Specifically, we are building \sysname, a system that supports in-DB model inference and leverages sophisticated cross-optimizations and tight integration of ML runtimes in the DB to outperform common practical solutions by up to $24\times$.

In our vision, data scientists should be able to design and train ML 
models with their favorite ML framework. Once trained, these models, 
combined with any required data preprocessing steps and library 
dependencies, form what we call a {\em model pipeline}. 
\sysname supports model pipelines expressed in a generic and 
portable model format \cite{mlflowmodel} that is compatible with 
MLflow \cite{mlflow}, and stores them in the RDBMS.
Users can then invoke them (on data stored in the DB or on fresh data coming from an application) 
by issuing SQL commands. We call a SQL query that invokes a model pipeline an~\textbf{inference query}.

Delivering competitive performance for in-DB inference requires a substantial engineering and research effort. Our running example touches upon several of the
interesting opportunities we discovered in doing so (\autoref{sec:archi}).
\sysname introduces an intermediate representation (IR) that includes both ML
and relational operators. Input inference queries are captured in this IR by means of static analysis (\autoref{sec:ir}). The IR is then analyzed and optimized using novel cross-operator optimizations and transformations that \sysname 
proposes (\autoref{sec:opt}). Finally, the optimized IR is fed for execution to the 
DBMS that supports different ML runtimes.
To achieve best-in-class performance for our optimized plans, we integrated ONNX 
Runtime\footnote{ONNX Runtime~\cite{ort} is a state-of-the-art inference engine 
with support for diverse environments and backends, which we built and open-sourced 
at Microsoft. It supports all models that can be expressed in ONNX~\cite{onnx}, \ie the vast 
majority of models.} natively within SQL Server (\autoref{sec:execution}).
Given that support for native execution of all ML pipelines is elusive, \sysname also employs out-of-process~\cite{spexternalsql} and containerized execution~\cite{sqlbdc} as required to achieve 100\% coverage.

We show that (i)~SQL Server with integrated ONNX Runtime is a solid 
building block for high-performance inference---yielding up to $5.5\times$ speedups over standalone solutions;
(ii)~\sysname's cross-optimizations yield benefits of up to $24\times$.

While \sysname is far from a finished system, the existing implementation already demonstrates the great potential for both research and industrial impact, by extending DBMSs with their robust capabilities to handle inference.  We are busy incorporating the techniques we present in this paper in a full-fledged cost-based optimizer---hardware acceleration and multi-query optimization will make this even more fun.

\section{\sysname Overview}
\label{sec:archi}

Our running example is {\em predicting the duration of stay in a hospital},\footnote{The example is based on~\cite{los-rserver}, with changes designed to showcase several \sysname optimizations.} depicted in \autoref{fig:runningexample}.

A data scientist has developed a decision tree model \verb|M| that predicts a patient's length of stay in a hospital, by combining \verb|patient_info| with results from \verb|blood_tests| and \verb|prenatal_tests|.
The model is trained over large amounts of data (\eg across all hospitals in an insurance network)
and is deployed/stored within the RDBMS.
At a later time, an analyst, employed by a specific hospital, issues a SQL query \verb|Q| to apply the model on local data in order to ``{\em find pregnant patients with a high likelihood of staying in the hospital for more than a week}'' and inform the medical staff.

By storing and scoring the model within the RDBMS, we inherit ease of access via SQL, and several desirable properties regarding updates to the deployed model: {\em transactionality} (a change to the model is handled as part of a transaction), {\em high-availability}, and {\em auditability}. To achieve good performance, \sysname employs several optimizations and then performs inference natively within the RDBMS invoking an ML runtime as an integral part of the database runtime.

The input inference query \verb|QM|, which includes both the SQL query \verb|Q| and the model pipeline \verb|M| (in Python here), is handled as follows.
First, \sysname's \emph{Static Analyzer} parses \verb|QM| and performs static analysis on the SQL and Python scripts.
The result is a DAG expressed in \sysname's unified IR (detailed in \autoref{sec:ir}), as shown in \autoref{fig:runningexample}.

The IR is fed to the \emph{Cross Optimizer}, which performs various optimizations (passing information between the data operators and the ML ones) and operator transformations, and determines which part of the IR will be executed by SQL Server and which by the integrated ML runtime (ONNX Runtime here).
The space of optimization is very rich. Some representative optimizations (discussed further in \autoref{sec:opt}) are:
\begin{itemize}[noitemsep,topsep=0pt,parsep=0pt,partopsep=0pt,leftmargin=10pt]
\item {\em predicate-based model pruning}: the condition \verb|pregnant=1| is pushed upward and into the decision tree, resulting in the right subtree being pruned.
\item {\em model-projection pushdown}: unused or zero-weight features can be projected-out early in the query plan---this is common due to model regularization or due to the above pruning (e.g., \verb|gender| is no longer used).
\item {\em model/query splitting}: the pruned model can be partitioned in a cheap model (for \verb|age<=35|) and a more complex one (for \verb|age>35|). Model and query are thus split in two branches and separately optimized.\footnote{This shares commonalities with model cascades~\cite{willump}.}
\item {\em model inlining}: small decision trees can be inlined thanks to SQL Server's recent UDF inlining feature~\cite{froid}.
\item {\em NN translation}: \raven can transform many classical ML models (\eg decision tree) and featurizers into equivalent neural networks (NN) to then leverage the highly optimized ONNX Runtime for batch scoring on CPU/GPU.
\item {\em standard DB optimizations}: such as predicate/projection pushdown and join elimination can be triggered---in the inlined left-branch we don't need to join with \verb|prenatal_| \verb|tests|, and \verb|bp>140| can be derived and pushed-down.
\item {\em compiler optimizations}: we implemented compiler-style optimizations such as constant-folding within ONNX Runtime---the \verb|pregnant| variable is a constant in our example query and can be propagated inside the NN.
\end{itemize}

\vspace{1mm}
The optimized \sysname IR is passed to the \emph{Runtime Code Generator}, which generates a new SQL query, reflecting the above optimizations. The integrated SQL Server+ONNX Runtime engine is then invoked for execution.\footnote{For inference queries that are not yet supported by our static analysis or by ONNX Runtime, we support calling external ML runtimes and containerized execution.}

It is clear from the above that extensive optimizations are possible once we bring ML inference into the DBMS. 
At the time of writing, we have added native support for ONNX Runtime within SQL Server. We have designed and implemented several of these optimizations, and automated the static analysis process. 
In the next sections, we describe the path we are taking towards building an optimizer and runtime for integrated evaluation of inference queries.

\section{\sysname IR and Static Analysis}
\label{sec:ir}

Intermediate representations, have been commonly used for enabling optimizations in various settings. Most database query optimizers rely on relational algebra, whereas different IRs have been proposed for ML runtimes~\cite{xla,onnx}. 

In \sysname, we chose to combine both data and ML operators in a \emph{unified} IR, as shown in \autoref{fig:runningexample}.
This allows us to optimize an inference query, which includes both data and ML operations, in a holistic manner:
we can perform optimizations that span data and ML operations, and pick the most suitable runtime to execute each operator (\autoref{sec:opt}).

Next, we define \sysname's IR and describe the static analysis process to extract the IR.

\subsection{\sysname IR}
\label{s:ir:ir}

\sysname's data and ML operators are chosen to cover most practical scenarios, based on our analysis of \mysim4.6 million publicly available Python notebooks from GitHub. 
Our current operator set, which is easily extensible, can be split into the following categories.

\mypar{Relational algebra (RA)} This includes all the relational algebra operators, which are found in a typical RDBMS.

\mypar{Linear Algebra (LA)} A large fraction of the operators used in ML frameworks, and in particular neural network runtimes~\cite{tensorflow,ort,pytorch}, fall into this category. Examples include \texttt{matrix multiplication} and \texttt{convolution} operators.

\mypar{Other ML operators and data featurizers (MLD)} These are operators widely used in classical (non-NN) ML frameworks (scikit-learn~\cite{sklearn}, ML.NET~\cite{mlnet}), but do not fall in the LA category, such as
decision trees and featurization operations (\eg categorical encoding, text featurization).

\mypar{UDFs} When the static analyzer is not able to map part of the input into operators of the above categories (e.g., a function containing arbitrary Python code), a UDF operator is used to wrap the non-optimizable code as a black box.

\vspace{1mm}
Note that our IR includes both higher- and lower-level operators. For example, a linear regression operator (higher-level) can also be expressed as a set of linear algebra operators (lower-level). We purposely allow diverse operator levels to unlock different optimizations, similar to MLIR~\cite{mlir}.

\subsection{Static Analysis}
\label{s:ir:static}

An inference query consumed by \sysname (see \autoref{fig:runningexample}) is a SQL query that performs (part of) the data processing and invokes ML model pipelines.\footnote{There is no standardized way yet to invoke models in SQL. Here we use the SQL Server way (as of version 2017) through the \texttt{PREDICT} or the \texttt{sp\_execute\_external\_script} statements~\cite{predictsql,spexternalsql}.}
The whole inference query can be instead expressed as a script in some imperative language (e.g., Python or R).
The input scripts are accompanied by metadata to specify the required runtimes and dependencies (e.g., Python version, libraries used), and to access the referenced data and models.
An open model format, such as the one defined in MLFlow~\cite{mlflowmodel}, can be used for this purpose.

Translating the SQL part into the IR is straightforward (similar to a DB parser that builds a logical plan). The interesting part is analyzing the model scripts expressed in an imperative language. Our current prototype supports Python scripts and notebooks (given their popularity in ML~\cite{kagglesurvey}).

Given a Python script, the Static Analyzer performs lexing, parsing, extraction of variables and their scopes, semantic analysis, type inference, and finally extraction of control and data flows. 
To compile the dataflow to an equivalent IR plan, the Static Analyzer takes as input an in-house knowledge base of APIs of popular data science libraries (e.g., Pandas~\cite{pandas}, NumPy~\cite{numpy}, scikit-learn~\cite{sklearn}, PyTorch~\cite{pytorch}), along with functions that map dataflow nodes/subgraphs to equivalent IR operators. 
Dataflow parts that cannot be translated to IR operators are translated to UDFs. 

This static analysis process comes with several challenges and limitations (again, we use UDFs when we cannot overcome them).
First, translating loops to relational or linear algebra operators is known to be a hard, if not undecidable, problem~\cite{ahmad:2018}. In our analysis of the \mysim4.6 millions Python notebooks, however, we found that only \mysim$17\%$ of all notebook code cells use such constructs. Thus, the vast majority of cases can be handled through analysis of straight line code blocks. 
Second, conditionals result in potentially multiple execution paths. In such cases, the Static Analyzer will extract one plan per execution path. 
Hence, downstream components in Raven need to operate based on multiple plans. 
Third, in dynamically typed languages, such as Python, type inference may result in assigning a collection of potential types to variables.
We plan to use knowledge from the SQL part to improve type inference in many practical~scenarios.

In most practical cases we tested, static analysis takes less than $10$\,msec. 
Its end result is a \sysname IR plan that is given as input to the Cross Optimizer, discussed next.

\begin{figure*}[t!]
	\centering\includegraphics[width=0.24\textwidth]{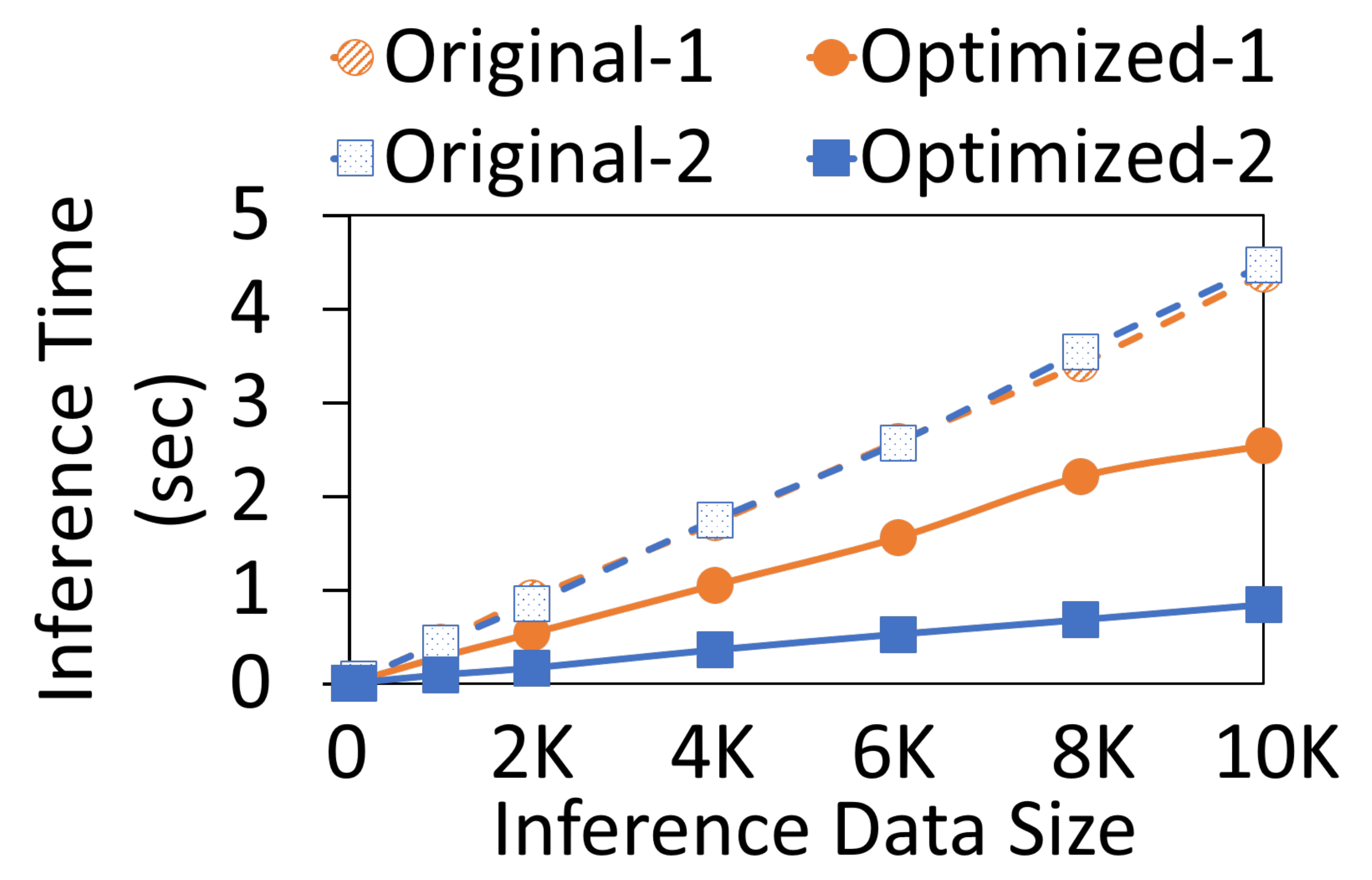}
	\centering\includegraphics[width=0.2\textwidth]{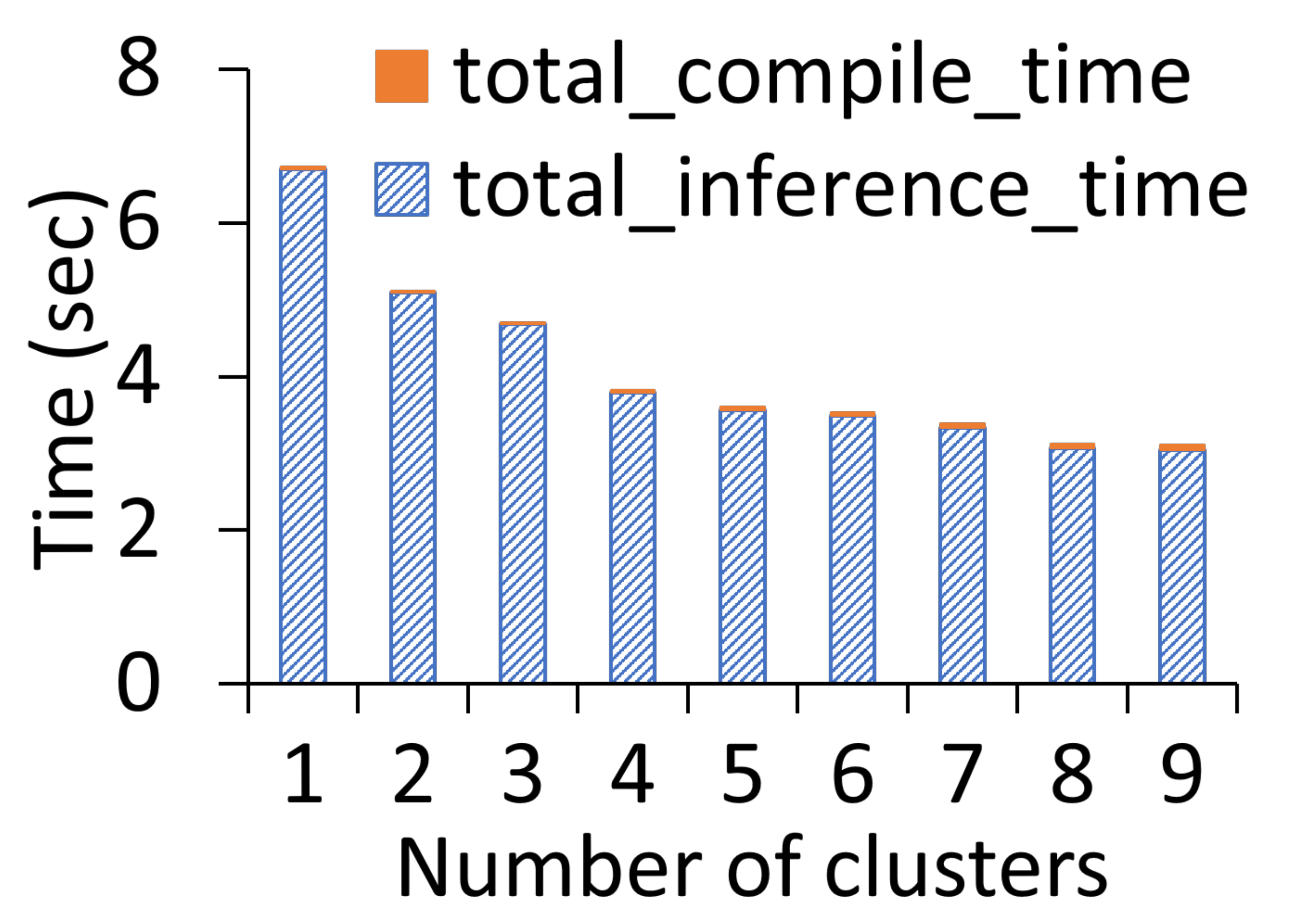}
	\centering\includegraphics[width=0.24\textwidth]{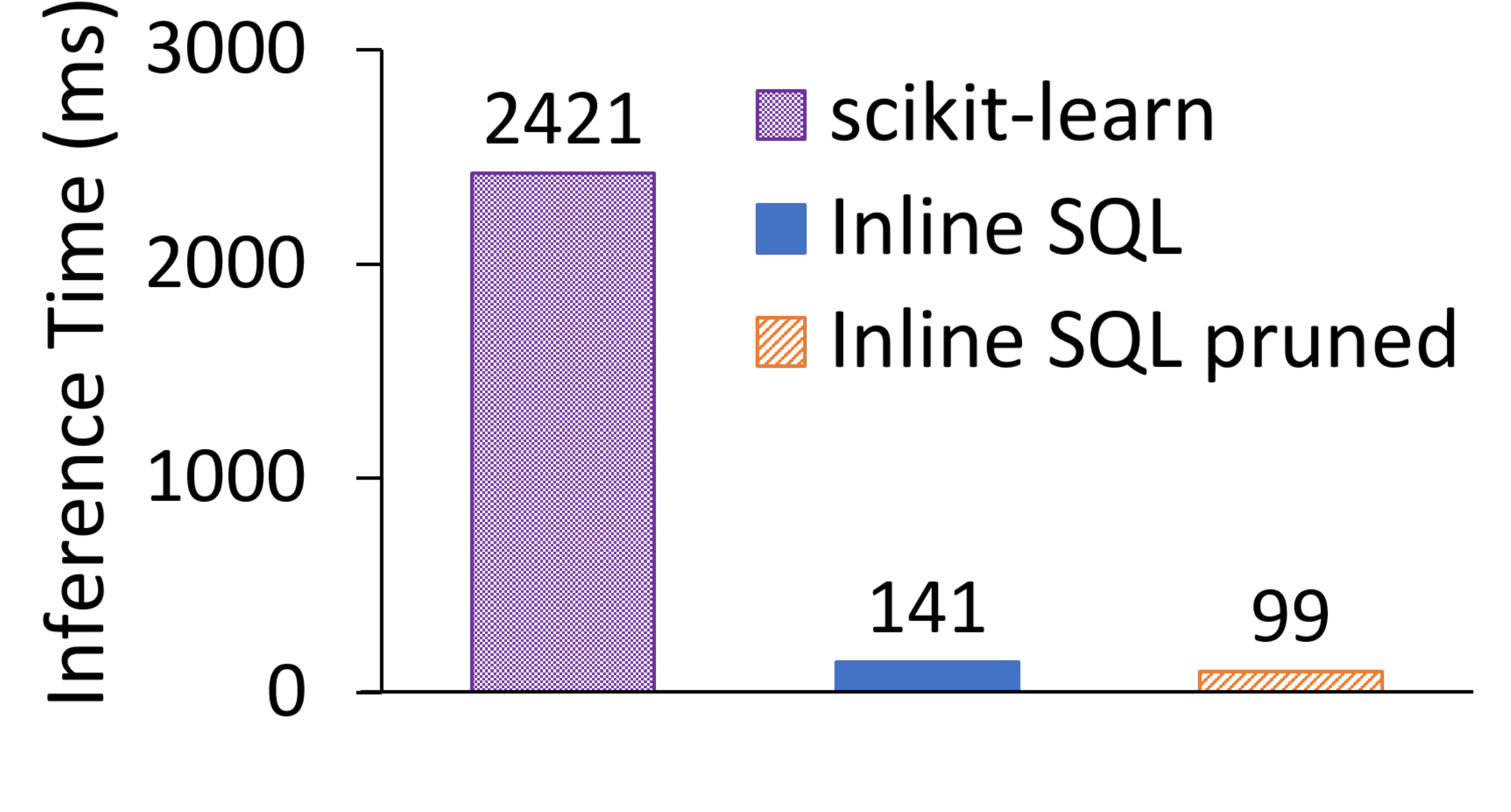}
	\centering\includegraphics[width=0.28\textwidth]{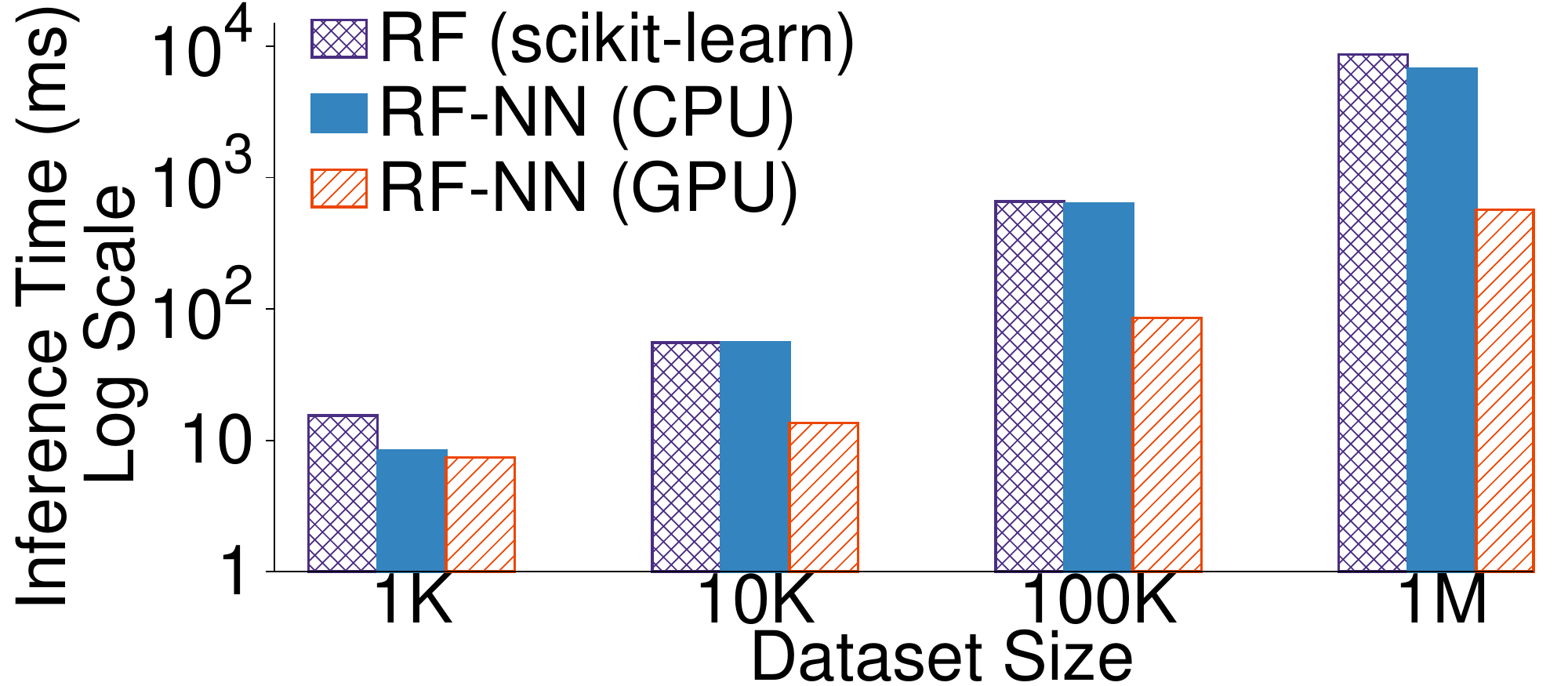}
	\caption{Left to right: cross-optimizations for flight delay ((a)~model-projection pushdown and (b)~model clustering) and operator transformations for hospital stay ((c)~model inlining and (d)~NN translation).\label{fig:all-opts-eval}}
\end{figure*}

\section{Cross-Optimizations}
\label{sec:opt}

In this section, we focus on the novelty aspects of our optimizer: cross-IR optimizations that pass information between data and ML operators (\autoref{sec:opt:crossir}), and transformations between operators to allow more efficient engines to be used for the same operation (\autoref{sec:opt:optransform}). All our optimizations can be expressed as transformation rules, applied by \sysname's Cross Optimizer (\autoref{sec:opt:opt}). By using state-of-the-art relational and ML engines, \sysname can also leverage the large body of work in relational and ML inference optimization~\cite{ort,tflopt,tvm,pretzel}---we do not further discuss such techniques here.

While discussing each optimization, we also evaluate its benefits, using two datasets:
(i)~patient information to predict length of stay in hospital (per our running example in \autoref{sec:archi});
(ii)~flight information to predict whether a flight will be delayed.\footnote{\url{https://www.kaggle.com/usdot/flight-delays}}
We use dataset sizes of up to 10M tuples for inference ($1.25$\,GB on disk).
Unless mentioned otherwise, all experiments are run on an Azure D16s\_v3 VM instance, with $16$\,vCPUs, $64$\,GB of RAM, and a $1.1$\,TB
SSD. 
Numbers are averages over multiple warm runs, and for each run we count the time it takes to load the model, perform the optimization, read the data, and perform inference over them.

\subsection{Cross-IR optimizations}
\label{sec:opt:crossir}

In this set of optimizations, we exploit ML operator characteristics (\eg model weights) to optimize data operations of the inference query (\emph{model-to-data}), or leverage relational operator- and data-properties to optimize the ML part of the query (\emph{data-to-model}).
Below we present some first such techniques we have devised---many more can be introduced.

These cross-IR optimizations can be seen as a form of Side-ways Information Passing (SIP)~\cite{sipRaghu}.
However, unlike SIP that requires adapting physical operators, our techniques are applied purely at query optimization time.

\vspace{1mm}
\mypar{Predicate-based model pruning} 
This data-to-model optimization exploits predicates in the IR (\eg coming from the \texttt{WHERE} clause of the SQL query or a Pandas' filter) to simplify a model.

In our running example (\autoref{fig:runningexample}), we can propagate the filter \texttt{pregnant=1} to the downstream decision tree model. 
The right branch of the tree can then be eliminated, thereby improving its prediction time---by $29\%$ in our example.

Predicate-based pruning can also be beneficial for categorical features. Such features are typically encoded as a set of binary features, one for each unique value of the original feature. If there is a selection on the original feature, only one of the corresponding binary features will be non-zero. Hence, the rest of the features can be dropped from the model.
We trained a logistic regression model for the flight delay and added a filter on the destination airport---predicate-based pruning yields a $\sim$2.1$\times$ on this query using scikit-learn, regardless of the filter's selectivity (what matters in this speed up is the number of features dropped).

Likewise, we can improve a neural network's performance via constant folding,\footnote{\url{https://github.com/microsoft/onnxruntime/blob/master/onnxruntime/core/optimizer/constant_folding.h}} \ie statically computing part of the model based on the constant input from the predicate.

This technique can also be applied based on data properties instead of explicit selections. Using data statistics,
we might observe that only specific unique values appear in the data or that data follows specific distributions (\eg all patients are above $35$). In these cases, we can derive predicates to perform predicate-based pruning.

\vspace{1mm}
\mypar{Model-projection pushdown} In this model-to-data optimization, we observe properties of an ML operator to simplify the data processing part of the inference query. 

Consider a logistic or linear regression model with some of its weights being zero. This is often the case when $L_1$-regularization techniques are applied during training (e.g. Lasso) to improve the model's generalization ability, size, and prediction cost. Here we exploit this property further. The features that will be multiplied with these zero-weights are not useful for the prediction, and can be projected out and removed from the model without affecting the inference result.

We trained logistic regression models for flight delay, using scikit-learn and various $L_1$-regularization strengths.\footnote{\url{https://scikit-learn.org/stable/modules/generated/sklearn.linear_model.LogisticRegression.html}} We picked the two highest-performing models (with highest AUC): the one had $41.75\%$ sparsity (i.e., percentage of zero weights), the other $80.96\%$. \autoref{fig:all-opts-eval}(a) shows that model-projection pushdown improves inference time by $\sim$1.7$\times$ for the first model and $\sim$5.3$\times$ for the~second.

Model-projection pushdown might be enabled by other optimizations: in \autoref{fig:runningexample}, predicate-based pruning of the right tree branch enables model-projection pushdown on \texttt{gender}, as it is no longer needed.
Similarly, it can enable other optimizations: after eliminating features, the relational optimizer can drop joins if one of the joining relations no longer provides features needed by the model.

There are several more questions we plan to investigate: What is the impact of physical database design, such as column stores, when applying model-projection pushdown? What is the benefit for more complex models, such as NNs? What would be the impact in runtime and model accuracy when applying \emph{lossy} model-projection pushdown, where small, but non-zero, weights are removed?

\mypar{Model clustering} Taking predicate-based pruning using data properties a step further, we may not have a single value for one or more features, but can cluster the data in a way that each cluster has specific values for some features. We can then precompile optimized models for each cluster.

We performed k-means clustering with an increasing number of clusters for $700$K tuples of flight delay.
\autoref{fig:all-opts-eval}(b) shows that model clustering reduces inference time by up to $54\%$. The more the clusters the bigger the gain, although the relative gain diminishes after a point.
Model compile time, \ie the time to create new models by dropping features, is negligible.
On the other hand, hospital stay does not benefit from clustering since its categorical features are already binary, therefore fewer features are dropped.

Clustering can be relatively expensive, depending on input data size ($0.4$ to $42$\,secs in our examples).
In practical settings, clustering can be performed offline on a sample of historical data. When new data arrives, if a precompiled model does not exist, we fall back to the original model.

\subsection{Operator Transformations}
\label{sec:opt:optransform}

Along with the logical optimizations presented above, \sysname applies rules that transform (a set of) operator(s) to another. For example, we can map a linear regression to a matrix multiplication. 

Such transformations enable both additional optimizations and the use of different runtimes for executing an operator (\eg a high-performance NN engine might not have a dedicated linear regression, but has the lower-level operator it got translated to).
Note that transformations should preserve semantics (\eg SQL's bag semantics vs. Pandas' ordered bags).

\mypar{Model inlining} These transformations translate ML operators (LA and MLD operators, see \autoref{sec:ir}) to relational ones. Several of these transformations have been studied in the literature~\cite{laradb,levelheaded,systemml,ml2sql}. They are particularly important in \sysname, because they allow us to use the relational optimizations and high performance of SQL Server for data operations (\eg a join that was initially expressed in Python).
Moreover, we employ the UDF inlining technique introduced in SQL Server 2019~\cite{froid} to further boost performance.

We trained a decision tree (the same technique would work for tree ensembles) for the hospital stay in scikit-learn, translated it to a UDF after expressing its conditions as a SQL query, and inlined the UDF in the query. \autoref{fig:all-opts-eval}(c) demonstrates that this ML-to-relational operator translation yields a performance gain of \mysim$17\times$ for a dataset of $300$K tuples when compared to running the decision tree in scikit-learn reading data from the DB (reading from a CSV was similar). Big part of this gain was due to completely avoiding data transformations by keeping execution inside the DB.
Assuming a query with a selection on a tree's dimension, as discussed above, we can further improve runtime by $29\%$ with predicate-based pruning, leading to a total improvement of $24.5\times$.

We also experimented with pushing categorical encodings to SQL Server. Our initial experiments show significant performance improvements when the number of resulting features is not too big, but further investigation is required to draw safe conclusions.

\mypar{NN translation} \sysname introduces novel transformations from MLD (see \autoref{sec:ir}) to linear algebra operators. 
This allows us to express classical ML operators and data featurizers, typically written in frameworks like scikit-learn and ML.NET, to neural-networks that can be executed in highly efficient engines like ONNX Runtime, PyTorch, and TensorFlow. This is very important performance-wise: unlike most traditional ML frameworks, NN engines support out-of-the-box hardware acceleration through GPUs/FPGAs/NPUs, as well as code generation~\cite{tvm}.

In \autoref{fig:all-opts-eval}(d), we compare a random forest model (RF) for hospital stay in scikit-learn against the NN translation of the same model (RF-NN), both on CPU and GPU.
Here we used a machine with similar specs to our VM but equipped with an Nvidia K80 GPU.
For $1$K tuples, RF-NN on CPU is about $2\times$ faster compared to the RF on scikit-learn, whereas RF-NN on GPU further decreases computation time by $10\%$. As we increase the dataset size, the gap between scikit-learn and RF-NN on CPU closes, and both have performance increasing almost linearly to the dataset size.
Conversely, with larger datasets we can better utilize the parallel architecture of the GPU, and therefore get a speed-up of up to $15\times$ compared to scikit-learn for $1$M tuples.

\subsection{\sysname's Cross Optimizer}
\label{sec:opt:opt}

So far, we discussed various transformation rules (cross optimizations and operator transformations) that we have introduced and implemented in \sysname, and showed their benefits on real models/data.
The next important step in our journey is to integrate all these rules in our optimizer---we are actively working on this at the time of writing.
An initial version will be heuristic-based, applying all rules in a specific order. Our goal is to then get to a cost-based Cascades-style optimizer, possibly integrated with SQL Server optimizer, in which each operator will be associated with a cost. Several plan alternatives will be considered by applying the rules in different orders and the best will be picked.
Note that as part of the optimization process, we need to pick the runtime that each operator will be executed in (relational engine or ML runtime), based on each runtime's capabilities and performance (including specialized hardware), and the cost of switching across engines.

\section{Inference Query Execution}
\label{sec:execution}

\sysname's Runtime Code Generator builds a new SQL query that corresponds to the optimized IR (\ie the output of the Cross Optimizer).
The model invocations that are included in the optimized SQL query will be executed in one of the following ways (in decreasing level of integration with SQL Server's main relational engine):

\mypar{In-process execution (\sysname)} Starting with version 2017, SQL Server introduced the \texttt{PREDICT} statement~\cite{predictsql} to allow native inference for a small set of models. As part of realizing our vision, we deeply integrated ONNX Runtime inside SQL Server. ONNX Runtime is used as a dynamically linked library to create inference sessions, transform data to tensors, and invoke in-process predictions over any ONNX model or any model that can be expressed in ONNX through \sysname's static analysis or ONNX converters~\cite{onnxconvert}.
A user simply needs to store their model in SQL Server and issue queries that include model inference using the existing \texttt{PREDICT} statement. This is the tightest-integration option: apart from in-process execution, it also allows us, out-of-the-box, to take advantage of model and inference-session caching, and SQL Server's optimizer.

\mypar{Out-of-process execution (\sysname Ext)} For model pipelines not yet supported by our Static Analyzer, we use the \texttt{sp\_execute\_external\_script}~\cite{spexternalsql} statement, which instantiates an external language runtime to perform out-of-process inference. Currently supported languages are Python and R (and Java starting with 2019 version).

\begin{figure}[t]
	\centering\includegraphics[width=0.87\columnwidth]{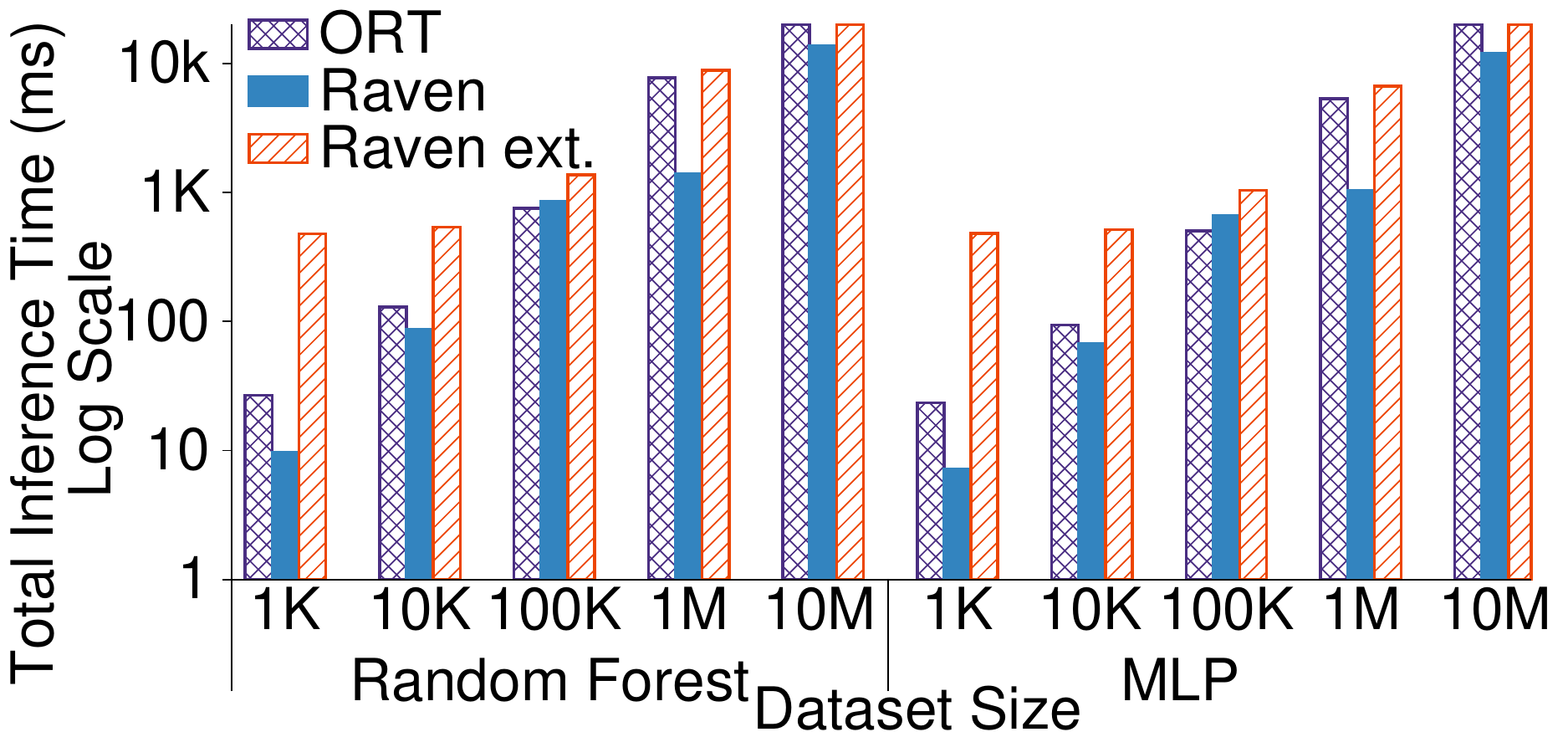}
	\caption{Model inference performance for SQL Server with in- and out-of-process ONNX Runtime (\sysname and \sysname Ext, resp.) and standalone ONNX Runtime (ORT).}
      \label{fig:sqls-ort} %
            \vspace{-2mm}
\end{figure}

\mypar{Containerized execution} For model pipelines that cannot be executed with one of the above techniques (\ie written in a language that is supported neither by our Static Analyzer nor by \texttt{sp\_execute\_external\_script}), we fall back to spinning up a Docker container and performing inference through a REST endpoint~\cite{sqlbdc}.

\vspace{1mm}
Having shown the substantial benefits of our optimizations in \autoref{sec:opt}, we turn to the following question: \emph{Can our in-process integration of ONNX Runtime with SQL Server match the performance of a standalone ONNX Runtime instance when performing pure model inference (no SQL part)?} Or are there significant overheads in the~integration?

We compare: (i)~standalone ONNX Runtime (\emph{ORT} hereafter), (ii)~\sysname, and (iii)~\sysname Ext. 
We use both a random forest (RF) and a multi-layer perceptron (MLP) as part of a pipeline that also includes featurization, and translate both end-to-end pipelines to NNs to be efficiently executed in ORT (see NN translation, \autoref{sec:opt}). 
\autoref{fig:sqls-ort} shows results for increasing dataset size.

We make the following observations:

\noindent (i)~Between 50K and 100K tuples, ORT and \sysname have similar performance, with \sysname having an overhead of up to 15\%. We are positive that we can further close this gap with implementation improvements. That said, this overhead is insignificant, compared to the orders-of-magnitude improvements that \sysname's optimizations~provide.\footnote{Some of our optimizations could be applied to an ML runtime too. However, they would lack the relational engine highly optimized for operations like joins and aggregates, and the other benefits of an RDBMS~\autoref{sec:intro}.}

\noindent (ii)~For smaller datasets (\eg up to 50K tuples) and warm runs, \sysname is faster than ORT (\eg 3msec vs. 20msec for 100 tuples). This is due to SQL Server's model and inference-session caching across queries, instead of loading the model from disk and relying on the file system cache.

\noindent (iii)~For 1M and 10M tuples \sysname is faster than ORT by around 5$\times$! This came to our surprise. After investigation, we observed that for larger datasets, SQL Server \emph{automatically} parallelizes both the scan and \texttt{PREDICT} operators. When forcing sequential execution, \sysname was about 7\% slower than ORT.
This model inference could potentially be parallelized in ORT as well, but that would not be trivial, whereas it came for free with SQL Server.

\noindent (iv)~\sysname Ext has a constant overhead of about half a second to start the external language runtime and some additional overheads, most probably due to data transfers. Still it is a viable option in cases our Static Analyzer does not support the model pipeline.

\noindent (v)~In our implementation of \sysname, we gained about an order-of-magnitude by performing batch inference instead of one prediction per tuple (ideal batch size to be investigated).

\section{Related Work}

Several previous works have proposed to run machine learning into the RDBMS~\cite{madlib,simsql2,systemml,levelheaded,doppiodb}. 
Interesting enough, these works mostly focus on training, whereas the prime focus of Raven is inference of already trained machine learning models. 
Other popular approaches for model inference~\cite{fromtheedgetothecloud} are containerized~\cite{clipper} and in-application~\cite{mlnet} execution.
As we argued in the introduction, model inference has interesting algebraic characteristics which makes it a likely (easier) target of integration with database query optimizer and runtime.
Cross-optimization of relational and linear algebra operators is recently becoming a hot topic~\cite{laradb,lara}, whereas specific optimizers~\cite{pretzel,willump} and runtimes~\cite{onnx,tvm} for model inferences are starting to emerge as well.
Our goal with Raven is to bridge the gap between the two worlds: we propose an optimizer able to execute both runtime-specific and cross-IR optimizations in an end-to-end~fashion.

\section{Conclusion}
\label{sec:future}

We presented \sysname, a system we are building  to
perform in-DB ML inference. \sysname performs static analysis of Python ML pipelines and SQL queries, that are captured in a unified IR. This enables us to apply novel cross-optimizations, yielding performance gains of up to $24\times$. The target execution environment for this optimized IR is a deeply integrated version of SQL Server and ONNX Runtime, which alone provides up to $5.5\times$ performance gains over
standalone ONNX Runtime execution. This is only the beginning of a long journey to incorporate ML scoring as a foundational extension of relational algebra, and an integral part of SQL query optimizers and runtimes.

\printbibliography

\end{document}